\documentclass[a4paper,fleqn,usenatbib]{mnras}
\usepackage{graphicx}
\usepackage{epstopdf}
\usepackage[T1]{fontenc}
\usepackage{ae,aecompl}

\usepackage{graphicx}	% Including figure files
\usepackage{amssymb}	% Extra maths symbols

\usepackage{savesym}
\usepackage{amsmath}
\savesymbol{iint}
\usepackage{txfonts}
\restoresymbol{TXF}{iint}

\usepackage{mathrsfs}

\def\({\left(}
\def\){\right)}
\def\[{\left[}
\def\]{\right]}
\def\e{\begin{equation}}
\def\q{\end{equation}}
\def\m{\begin{eqnarray}}
\def\n{\end{eqnarray}}

\title{Finsler space-time can explain both parity asymmetry and power deficit seen in CMB temperature anisotropies}

\author[Zhe Chang et al.]{
Zhe Chang,$^{1,2}$
Pranati K. Rath,$^{1,3,4}$
Yu Sang,$^{1,2}$
Dong Zhao$^{1,2}$\thanks{E-mail: zhaod@ihep.ac.cn}
and Yong Zhou$^{1,2}$
\\
$^{1}$Institute of High Energy Physics, Chinese Academy of Sciences, Beijing 100049, China\\
$^{2}$University of Chinese Academy of Sciences, Beijing 100049, China\\
$^{3}$Theoretical Physics Center for Science Facilities, Chinese Academy of Sciences, Beijing 100049, China\\
$^{4}$Khallikote Autonomous College, Berhampur 760001, India
}

\date{Accepted XXX. Received YYY; in original form ZZZ}

\pubyear{2018}

\begin{document}
\label{firstpage}
\pagerange{\pageref{firstpage}--\pageref{lastpage}}
\maketitle

\begin{abstract}
We propose a framework of Finsler space-time to explain the observed parity asymmetry and the power deficit in the
low-$\ell$ ($2\leqslant \ell \leqslant 29$) multipole range of cosmic microwave background (CMB) temperature anisotropies.
In the $3+1$ dimensional space-time, the three-dimensional space is described by a Randers-Finsler space, which is spatially irreversible,
inducing the parity asymmetry in the CMB angular power spectrum. We estimate the constraints on the two parameters introduced by Finsler space-time
via analyzing the low-$\ell$ angular power spectrum in PLANCK 2015 CMB temperature data. We see that the low-$\ell$ power
suppression in the CMB temperature anisotropies can also be resolved in this scenario. Our study shows that the two low-$\ell$ anomalies,
\emph{i.e.,} parity asymmetry and power deficit, may have a common origin.
\end{abstract}


%%%%%%%%%%%%%%%%%%%%%%%%%%%%%%%%%%%%%%%%%%%%%%%%%%

%%%%%%%%%%%%%%%%% BODY OF PAPER %%%%%%%%%%%%%%%%%%

\section{Introduction}\label{sec:introduction}
The cosmological principle says that the Universe is statistically homogeneous and isotropic on large scales.
Based on it, a standard cosmological model has been well established, all the six base parameters of which have been
constrained to a few percent level via performing data analysis of the Cosmic Microwave Background (CMB) temperature anisotropies and polarizations
\citep{Bennett:2012zja,Ade:2015xua}. However, several large-scale \emph{anomalies} has been reported by WMAP \citep{Hinshaw:2012aka} and PLANCK \citep{Ade:2015hxq} team.
For the CMB temperature anisotropies, the anomalies include the power suppression \citep{Efstathiou:2003hk,Bonga:2015kaa,Cai:2015xla,Chang:2011jw,Zhao:2014kca,Chang:2018msh},
the parity asymmetry \citep{Hansen:2012qd,Aluri:2011wv,Liu:2013wfa,Zhao:2013jya,Land:2005jq,Kim:2010gf,Kim:2010gd,Gruppuso:2010nd},
an alignment of quadrupole and octopole \citep{Chang:2013lxa,Copi:2013jna,Land:2005ad,Copi:2003kt,Abramo:2006gw},
the hemispherical power asymmetry \citep{Chang:2013vla,Rath:2013yra,Rath:2014cka,Jain:2014cpa,Eriksen:2007pc}, and
the lack of angular power on angular scales larger than 6$0^{\circ}$ \citep{Spergel:2003cb,Copi:2008hw,Copi:2013cya}.

Among these anomalies, the first two are related with this study. The power suppression of the CMB temperature anisotropies shows that the low-$l$ angular power spectrum,
especially the quadrupole, was found to be lower than what predicted by the standard cosmological model \citep{Hinshaw:1996uq,Hinshaw:2012aka,Adam:2015rua}.
The PLANCK data confirmed the power suppression of the CMB temperature anisotropies with 2$\sigma$-3$\sigma$ confidence level (CL)
in the multipole range $2\leq l\leq29$ \citep{Aghanim:2015xee}.
The parity asymmetry implies that the odd multipoles of CMB has excess power compared to the even multipoles.
In refs. \citet{Kim:2010gf,Kim:2010gd}, the authors investigated WMAP 7-year temperature data, and found odd parity preferences at low multipoles ($l\leq22$) with high
statistical significance (99.6\% CL).
PLANCK 2015 results confirmed the parity asymmetry with 2$\sigma$ CL in the mutipole range $l<20$ and with 3$\sigma$ CL in the mutipole range $20\leqslant l\leqslant30$ \citep{Ade:2015hxq}.

The connections between these anomalies are still open questions, and the physics behind them is worthy of study.
Several works towards this direction have been done, and three examples are as follows.
The study in ref. \citet{Zhao:2013jya} suggested that the parity asymmetry and the alignment of quadrupole and octopole may stem from the same source.
In ref. \citet{Kim:2010gf}, the authors showed that the low quadrupole anomaly may be correlated with the parity asymmetry, and the low quadrupole anomaly is part of the parity asymmetry rather than an independent anomaly.
The ref. \citet{Chang:2013lxa} showed that the quadrupole-octopole alignment may be related with the  hemispherical power asymmetry.

In this paper, we will propose a new approach to explain the parity asymmetry in the framework of Finsler space-time, and show that the power suppression of the CMB temperature anisotropies can be also resolved in this model.
As a generalization of Riemann geometry, Finsler geometry does not have the quadratic restriction on the metric \citep{Chern:2000}. Finsler spacetime admits less symmetries than the Riemann one does, and thus privileged directions can exist in Finsler spacetime \citep{Li:2010wv}.
In this work, we adopt the 3+1 dimensional Finsler spacetime, whose spatial part is described by a Randers-Finsler space \citep{Randers:1941gge}.
The three dimensional space is thus irreversible under the parity transformation.
This property would induce the parity asymmetry in the CMB temperature anisotropies.
In the low-$l$ range $2\leq l \leq 29$, the power of odd multipoles remains unchanged while the power of even ones are suppressed.
As an additional result, we will show that the power of low-$l$ multipoles in the CMB temperature anisotropies is suppressed in the Finslerian model.

The rest of the paper is organized as follows. In Section 2, we briefly review the Finsler geometry and the derivation of comoving curvature perturbation in Finsler space-time. In Section 3, we calculate the angular power spectrum of CMB temperature anisotropies in Finsler space-time
and introduce two new Finslerian parameters. As expected, the obtained angular power spectrum has an obvious odd-multipole preference.
In Section 4, we use the chi-square statistic of the parity asymmetry estimator to infer the constraints on the Finslerian parameters
using PLANCK CMB temperature data.
We also study the power suppression in the CMB temperature anisotropies using this model. Conclusions are given in Section 5.

\section{The comoving curvature perturbation in Finsler space-time}\label{sec:equation}\noindent
%In Riemann space, the line element is given by,
%\begin{equation}
%ds^2=g_{\mu\nu}(x)dx^{\mu}dx^{\nu}.
%\end{equation}
Finsler geometry is a natural generalization of Riemann geometry in absence of quadratic restriction \citep{Chern:2000}.
The foundation of Finsler geometry is the Finsler structure, denoted by $F(x,y)$, which is defined on the tangent bundle of a manifold $M$,
satisfying $F(x,\lambda y)=\lambda F(x,y)$ for any $\lambda >$ 0. Here $x$ denotes a position on $M$, and $y=dx/d\tau$ denotes a velocity.
The Finslerian metric is defined as
\begin{equation}
g_{\mu\nu}=\frac{\partial}{\partial y^{\mu}}\frac{\partial}{\partial y^{\nu}}\left(\frac{F^2}{2}\right)\ .
\end{equation}
The Finsler metric and its inverse are used to lower and raise the space-time indices.
%In Finsler space-time, the geodesic equation is given by the first-order variation of Finslerian length. It is given as
The geodesic spray coefficients $G^{\mu}$ are given as
%\begin{equation}
%$\frac{d^2x^{\mu}}{d\tau^2}+2G^{\mu}=0$,
%\ ,
%\end{equation}
%where $G^{\mu}$ is the geodesic spray coefficients taking the form
\begin{equation}
\label{gsc}
G^{\mu}=\frac{1}{4}g^{\mu\nu}\left(\frac{\partial^2F^2}{\partial x^{\lambda}\partial y^{\nu}}y^{\lambda}-\frac{\partial F^2}{\partial x^{\nu}}\right)\ .
\end{equation}
%Substituting Eq.(\ref{Fsp}) into Eq.(\ref{gsc}), we obtain
%\begin{eqnarray}
%\label{Gt}
%G^t&=&\frac{1}{2}\dot\Psi y^ty^t+\Psi_{,i}y^ty^i+\frac{1}{2}\left[a\dot a(1+2\Phi-2\Psi)+a^2\dot\Phi\right]F^2_{Ra}\ ,\\
%\label{Gi}
%G^i&=&\left(H+\dot\Phi\right)y^ty^i+\Phi_{,k}y^ky^i+\frac{1}{2a^2}\Psi_{,j}\overline g^{ij}y^ty^t-\frac{1}{2}\overline g^{ij}\Phi_{,j}F^2_{Ra}+\frac{1}{2}\overline g^{ij}\overline g_{jk,m}y^my^k-\frac{1}{2}\overline g^{ij}F_{Ra}(\textbf{b}\cdot\textbf{y})_{,j}\ ,
%\end{eqnarray}
%where $\overline g^{ij}$ represents the Finslerian metric in the Rander space $F_{Ra}$. For simplicity, the direction of \textbf{y} is taken to be paralleled with the wave vector \textbf{k} in the momentum space \cite{Li:2015sja,Li:2015kua,Li:2017vuc}.
%Using the geodesic spray coefficients,
The Ricci scalar in Finsler geometry can be expressed in terms of $G^{\mu}$, namely \citep{Chern:2000}
\begin{equation}
\label{sric}
Ric\equiv\frac{1}{F^2}\left(2\frac{\partial G^{\mu}}{\partial x^{\mu}}-y^{\lambda}\frac{\partial^2G^{\mu}}{\partial x^{\lambda}\partial y^{\mu}}+2G^{\lambda}\frac{\partial^2G^{\mu}}{\partial y^{\lambda}\partial y^{\mu}}-\frac{\partial G^{\mu}}{\partial y^{\lambda}}\frac{\partial G^{\lambda}}{\partial y^{\mu}}\right)\ .
\end{equation}
Correspondingly, the Ricci tensor $Ric_{\mu\nu}$ and the scalar curvature $S$ are, respectively, defined as \citep{HA:z}
\begin{eqnarray}
Ric_{\mu\nu}&=&\frac{\partial^2(\frac{1}{2}F^2Ric)}{\partial y^{\mu}\partial y^{\nu}}\ ,\\
\label{tric} S&=&g^{\mu\nu}Ric_{\mu\nu}\ .
\end{eqnarray}
In Finsler space-time, the gravitational field equation takes the form \citep{Li:2014taa,Li:2015sja,Li:2017vuc}
\begin{equation}
\label{gfe}
Ric^{\mu}_{\nu}-\frac{1}{2}\delta^{\mu}_{\nu}S=8\pi GT^{\mu}_{\nu}\ ,
\end{equation}
where $T^{\mu}_{\nu}$ is the energy-momentum tensor.
%In Finsler space-time, t
Finsler space-time is fully described by $F$. The Finslerian metric is reduced to a Riemannian metric if and only if $F^2$ is quadratic in $y$.

As a special case of Finsler geometry, Randers space has been used to study the anisotropic inflation \citep{Li:2015sja,Li:2015kua,Li:2017vuc}
and the anisotropic accelerating expansion of the universe \citep{Li:2015uda}. For the background space-time, the Finsler structure in this work is given by
\begin{equation}
\label{f2}
F^2=y^ty^t-a^2(t)F^2_{Ra}\ ,
\end{equation}
where $F_{Ra}$ denotes the Finsler structure of Randers space \citep{Randers:1941gge} and is given as
\begin{equation}
\label{fra}
F_{Ra}=\sqrt{\delta_{ij}y^iy^j}+\delta_{ij}b^iy^j\ .
\end{equation}
Here, for simplicity the spatial vector $\textbf{b}$ is of the form $b^i=\{0,0,b(\textbf{x})\}$ and $b(\textbf{x})$ depends only on the spatial coordinates $\textbf{x}$.
%For simplicity one can define an inner product as $\textbf{b}\cdot\textbf{y}=\delta_{ij}b^iy^j$.
In this work, we consider the slow-roll inflation. The action of the inflaton field is given by
\begin{equation}
S=\int d^4x\sqrt{-g}\left(\frac{1}{2}g^{\mu\nu}\partial_\mu\phi\partial_\nu\phi-V(\phi)\right),
\end{equation}
where $g=-a^3F^3_{Ra}/(\delta_{ij}y^iy^j)^{3/2}$ \citep{Chang:2013vla,Li:2015sja,Li:2017vuc}. The equation of motion for the first order perturbation of inflaton field is given as \citep{Li:2015sja}
\begin{equation}
\label{deltaphi1}
\delta\ddot{\phi}+3H\delta\dot{\phi}-a^{-2}\bar{g}^{ij}\partial_i\partial_j\delta{\phi}=0,
\end{equation}
where $H=\dot{a}/{a}$, $\bar{g}^{ij}$ represents the Finslerian metric of the Rander space and $\delta\phi$ is the first order perturbation of inflaton field. For simplicity, the direction of y is taken to be parallel with the wave vector $\textbf{k}$ in the momentum space \citep{Li:2015sja,Li:2015kua,Li:2017vuc}. Expanding $\delta\phi$ in Fourier modes, we can write the Eq.(\ref{deltaphi1}) as
\begin{equation}
\label{deltaphi2}
\delta\ddot{\phi}_{\textbf{k}_e}+3H\delta\dot{\phi}_{\textbf{k}_e}+a^{-2}k^2_e\delta\phi_{\textbf{k}_e}=0,
\end{equation}
where $\delta\phi_{\textbf{k}_e}$ represents the fourier expansion of $\delta\phi$ in Finsler spacetime and $k^2_e$ is the effective wavenumber, given by
\begin{equation}
k^2_e=\bar{g}^{ij}k_ik_j=k^2(1+b(\textbf{k})\hat{\textbf{k}}\cdot\hat{\textbf{n}}_z)^2.
\end{equation}
Working in conformal time $d\tau=dt/a$, we can obtain an exact solution of Eq.(\ref{deltaphi2}) as follows
\begin{equation}
\label{deltaphi3}
\delta\phi_{\textbf{k}_e}=\frac{e^{-ik_e\tau}}{a\sqrt{2k_e}}\left(1-\frac{i}{k_e\tau}\right)\simeq\delta\phi_{\textbf{k}}-i\frac{3H}{2\sqrt{2}k^{\frac{3}{2}}}e^{-ik\tau}b(\textbf{k})(\hat{\textbf{k}}\cdot\hat{\textbf{n}}_z),
\end{equation}
where $a=-1/H\tau$ and $\delta\phi_{\textbf{k}}$ represents the fourier modes of $\delta\phi$ in the standard inflation model, taking the form \citep{Riotto:2002yw}
\begin{equation}
\delta\phi_{\textbf{k}}=\frac{e^{-ik\tau}}{a\sqrt{2k}}\left(1-\frac{i}{k\tau}\right).
\end{equation}
The comoving curvature perturbation in Finsler spacetime is given as \citep{Li:2015sja,Li:2017vuc}
\begin{equation}
\label{RF}
\mathcal{R}_F=H\frac{\delta\phi_{\textbf{k}_e}}{\dot{\phi}_0}-\Phi,
\end{equation}
where $\Phi$ is the scalar perturbation of the Finsler structure F. Plugging the Eq.(\ref{deltaphi3}) into Eq.(\ref{RF}), we obtain
\begin{eqnarray}
\label{RF1}
\nonumber \mathcal{R}_F&=&H\frac{\delta\phi_{\textbf{k}}}{\dot{\phi}_0}-\Phi-i\frac{3H^2}{2\sqrt{2}\dot{\phi}_0k^{\frac{3}{2}}}e^{-ik\tau}b(\textbf{k})(\hat{\textbf{k}}\cdot\hat{\textbf{n}}_z)\\
&=&\mathcal{R}-A(k,\tau)b(\textbf{k})(\hat{\textbf{k}}\cdot\hat{\textbf{n}}_z),
\end{eqnarray}
where $\mathcal{R}=H\frac{\delta\phi_{\textbf{k}}}{\dot{\phi}_0}-\Phi$ denotes the comoving curvature perturbation in the standard inflation model and $A(k,\tau)=i\frac{3H^2}{2\sqrt{2}\dot{\phi}_0k^{\frac{3}{2}}}e^{-ik\tau}$. In Finslerian space-time, the extra part $A(k,\tau) b(\textbf{k})(\hat{\textbf{k}}\cdot\hat{\textbf{n}}_z)$ will induce anisotropic terms in the comoving curvature perturbation.
In next section, we will show that the anisotropic terms can explain the parity asymmetry and the power suppression of the CMB temperature anisotropies.

\section{The angular power spectrum of temperature anisotropies in the CMB}\label{sec:cmb}\noindent
The spherical harmonic coefficients of the CMB temperature fluctuation are given as
\begin{eqnarray}
\label{alm}
a_{lm}&=&4\pi(-i)^l\int\frac{d^3 \textbf{k}}{(2\pi)^{3/2}}\mathcal{R}_F\Delta_l(k)Y^*_{lm}(\hat{\textbf{k}})\nonumber\\
\label{alm2f}
&=&4\pi(-i)^l\int\frac{d^3\textbf{k}}{(2\pi)^{3/2}}\mathcal{R}\Delta_l(k)Y^*_{lm}(\hat{\textbf{k}})\nonumber\\
&&-4\pi(-i)^l\int\frac{d^3\textbf{k}}{(2\pi)^{3/2}}A(k,\tau) \nonumber b(\textbf{k})(\hat{\textbf{k}}\cdot\hat{\textbf{n}}_z)\Delta_l(k)Y^*_{lm}(\hat{\textbf{k}}),\\
\end{eqnarray}
where $\mathcal{R}_F$ and $\mathcal{R}$ represent the comoving curvature perturbation in Finsler spacetime and standard inflation model, respectively, and we have used the Eq.(\ref{RF1}) to get the second equality.
The first term at the right hand side of Eq.(\ref{alm2f}) corresponds to the usual FRW space-time, and its two-point correlation leads to the isotropic angular power spcetrum of CMB temperature anisotropies, whereas the second term is the correction term due to the Finsler space-time. %The two-point correlation of the second term
In fact, the Finslerian correction term can be represented as
\begin{eqnarray}
\label{almf}
(a_{lm})_F&=&-\frac{4\pi(-i)^l}{(2\pi)^{3/2}}\int_0^\infty dk\int_0^\pi d\theta_{\textbf{k}}\sin\theta_{\textbf{k}}\nonumber\\
&&\times\left[\int_0^{\pi}d\phi_{\textbf{k}}A(k,\tau) b(\textbf{k})(\hat{\textbf{k}}\cdot\hat{\textbf{n}}_z)\Delta_l(k)Y^*_{lm}(\hat{\textbf{k}})\right.\nonumber\\
&&\left.+\int_{\pi}^{2\pi}d\phi_{\textbf{k}}
A(k,\tau) b(\textbf{k})(\hat{\textbf{k}}\cdot\hat{\textbf{n}}_z)\Delta_l(k)Y^*_{lm}(\hat{\textbf{k}})\right] \nonumber\\
\nonumber &=&-\frac{4\pi(-i)^l}{(2\pi)^{3/2}}\int_0^\infty dk\int_0^\pi d\theta_{\textbf{k}}\sin\theta_{\textbf{k}}\nonumber\\
&&\times\left[\int_0^{\pi}d\phi_{\textbf{k}}A(k,\tau)b(\textbf{k})(\hat{\textbf{k}}\cdot\hat{\textbf{n}}_z)\Delta_l(k)Y^*_{lm}(\hat{\textbf{k}})\right.\nonumber\\
%&&~~~~~~~~~~~~~~~~~~~~~~~~~~~~~~~~~~~~~~~~~
\nonumber&&\left.+\int_0^{\pi}d\phi_{\textbf{k}}A(k,\tau)b(\textbf{-k})(\hat{\textbf{k}}\cdot\hat{\textbf{n}}_z)\Delta_l(k)(-1)^{l+1}Y^*_{lm}(\hat{\textbf{k}})\right].\\
\label{calalmf}
\end{eqnarray}
To obtain the second equality in the above derivation, the polar coordinates $(\theta, \phi)$ change to $(\theta'=\pi-\theta$, $\phi'=\phi-\pi)$, and the wave vector \textbf{k} changes to -\textbf{k} for the second term at the right hand side.
If the $b(\textbf{k})$ in Eq.(\ref{calalmf}) satisfies $b(-\textbf{k})=-b(\textbf{k})$, which is an ansatz of this work, the $(a_{lm})_F$ has an obvious odd-multipole preference. In this case, the Finslerian correction term reduces to
\begin{eqnarray}
\label{almF}
(a_{lm})_F&=&-\frac{4\pi(-i)^l}{(2\pi)^{3/2}}\int_0^\infty dk\int_0^\pi d\theta_{\textbf{k}}\sin\theta_{\textbf{k}}\nonumber\\
&&\times\int_0^{\pi}d\phi_{\textbf{k}}~[1+(-1)^{l}]A(k,\tau)b(\textbf{k})(\hat{\textbf{k}}\cdot\hat{\textbf{n}}_z)\Delta_l(k)Y^*_{lm}(\hat{\textbf{k}})\nonumber\\
\nonumber&=&-4\pi(-i)^l\int\frac{d^3\textbf{k}}{(2\pi)^{3/2}}\frac{[1+(-1)^l]}{2}A(k,\tau)\\
&&\times b(\textbf{k})(\hat{\textbf{k}}\cdot\hat{\textbf{n}}_z)\Delta_l(k)Y_{lm}^\ast(\hat{\textbf{k}}).
\end{eqnarray}
The equation shows that the Finslerian correction term works only for the even multipoles, while it always vanishes for the odd multipoles. %To be specific, we finally obtain
%\begin{eqnarray}
%(a_{lm})_F&=&
%\begin{cases}
%0& \text{for odd $\ell$}\ ,\\
%-4\pi(-i)^l\int\frac{d^3\textbf{k}}{(2\pi)^3}\frac{3}{2}(\textbf{b}\cdot\hat{\textbf{k}})\Delta_l(k)Y^*_{lm}(\hat{\textbf{k}})& \text{for even $\ell$}\ .
%\end{cases}%\\
%&=&
%-4\pi(-i)^l\int\frac{d^3\textbf{k}}{(2\pi)^3}\frac{3}{2}(\textbf{b}\cdot\hat{\textbf{k}})\Delta_l(k)Y^*_{lm}(\hat{\textbf{k}})\frac{[1+(-1)^{l}]}{2}\ .
%\end{eqnarray}
Therefore, we can rewrite the Eq.(\ref{alm2f}) as follows
\begin{equation}
\label{almFf}
a_{lm}=4\pi(-i)^l\int\frac{d^3\textbf{k}}{(2\pi)^{3/2}}[\mathcal{R}-\frac{[1+(-1)^{l}]}{2}A(k,\tau)b(\textbf{k})(\hat{\textbf{k}}\cdot\hat{\textbf{n}}_z)]\Delta_l(k)Y^*_{lm}(\hat{\textbf{k}})\ .
\end{equation}
Given the expression of spherical harmonic coefficients $a_{lm}$, the correlation of two spherical harmonic coefficients, namely $C_{ll'mm'}\equiv\langle a_{lm}a^*_{l'm'}\rangle$, can be given as
\begin{eqnarray}
\nonumber &&C_{ll'mm'}=4\pi\int\frac{dk}{k}P_R(\vec k)\Delta_l(k)\Delta^*_{l}(k)\delta_{ll'}\delta_{mm'}\nonumber\\
&&+16\pi^2(-1)^l(i)^{l+l'}\int\frac{d^3\textbf{k}}{(2\pi)^{3/2}}\int\frac{d^3\textbf{k}'}{(2\pi)^{3/2}}\nonumber\\
&&\times\left[\frac{[1+(-1)^{l}]}{2}\langle[A(k,\tau)b(\textbf{k})(\hat{\textbf{k}}\cdot\hat{\textbf{n}}_z)]\mathcal{R}^*(\textbf{k}')\rangle\right.\nonumber\\
&&\left.+\frac{[1+(-1)^{l'}]}{2}\langle[A(k',\tau)b(\textbf{k}')(\hat{\textbf{k}'}\cdot\hat{\textbf{n}}_z)]^*\mathcal{R}(\textbf{k})\rangle\right.\nonumber\\
&&\left.+\frac{[1+(-1)^{l}]}{2}\frac{[1+(-1)^{l^\prime}]}{2}\langle[A(k,\tau)b(\textbf{k})(\hat{\textbf{k}}\cdot\hat{\textbf{n}}_z)][A(k',\tau)b(\textbf{k}')(\hat{\textbf{k}'}\cdot\hat{\textbf{n}}_z)]^*\rangle\right]\nonumber\\
&&\times\Delta_l(k)\Delta^*_{l'}(k')Y^*_{lm}(\hat k)Y_{l'm'}(\hat {k'}),
\label{eq:xxxxx}
\end{eqnarray}
where $P_R(k)$ denotes the isotropic power spectrum of comoving curvature perturbation $\mathcal{R}$ in standard inflation model. In Finsler spacetime, the cross term of $\mathcal{R}$ and $b(\textbf{k})(\hat{\textbf{k}}\cdot\hat{\textbf{n}}_z)$ can generate anisotropic effect in the primordial power spectrum. As discussed in Ref. \citet{Li:2015sja,Li:2017vuc}, the anisotropic effect induced by the cross term only contributes to the off-diagonal part of $C_{ll'}$ ($l'=l+1$). In this work, we mainly consider the effect of quadrupolar modulation on the diagonal part of $C_{ll'}$ ($l'=l$), which is generated by the correlation between $b(\textbf{k})(\hat{\textbf{k}}\cdot\hat{\textbf{n}}_z)$ and $b(\textbf{k}')(\hat{\textbf{k}'}\cdot\hat{\textbf{n}}_z)$.
%Here, $P_R(k)$ is defined as we only considered a quadrupolar modulation quadrupolar modulation term
%is the primordial power specturm of the comoving curvature perturbation and obtained by using the definition
%\begin{equation}
%\langle\Phi(\textbf{k})\Phi^*(\textbf{k}')\rangle\equiv\frac{2\pi^2}{k^3}P_R(k)\delta^{(3)}(\textbf{k}-\textbf{k}')\ .
%\end{equation}
Here $P_R(k)$ takes the nearly scale-invariant form as
%\begin{equation}
$P_R(k)=A_s\left({k}/{k_0}\right)^{n_s-1}$,
%\end{equation}
where $A_s$ and $n_s$ denote the amplitude and spectral index, respectively. We set a pivot scale as $k_0=0.05\textrm{Mpc}^{-1}$ throughout this work.
%The second term at the right hand side of the Eq.(\ref{eq:xxxxx}) generates both off-diagonal ($l^\prime\neq l$) and diagonal ($l^\prime=l$) correlations.

As an ansatz, we assume a simple form of the correlation between $b(\textbf{k})(\hat{\textbf{k}}\cdot\hat{\textbf{n}}_z)$ and $b(\textbf{k}')(\hat{\textbf{k}'}\cdot\hat{\textbf{n}}_z)$ as follows
\begin{equation}
\label{bk}
\langle[A(k,\tau)b(\textbf{k})(\hat{\textbf{k}}\cdot\hat{\textbf{n}}_z)][A(k',\tau)b(\textbf{k}')(\hat{\textbf{k}'}\cdot\hat{\textbf{n}}_z)]^*\rangle=\frac{2\pi^2}{k^3}f(k)\delta^3(\textbf{k}-\textbf{k}')\ ,
\end{equation}
where the term $f(k)$ is a dimensionless quantity and depends only on the $k$. The effect of $b(\textbf{k})$ can be regarded as the first-order correction to the power spectrum in the $\Lambda$CDM model. It is possible to derive the negative spectra from the fluctuations in quantum field theory \citep{Hsiang:2010uv}. The contribution of the vacuum fluctuation in Finsler space-time is sub-dominant so that $f(k)$ in Eq.(\ref{bk}) could be negative.

The angular power spectrum of CMB temperature anisotropies in Finsler space-time can be written as
\begin{equation}
\label{cl}
C_{l}=4\pi\int\frac{dk}{k}\(P_R(k)+\frac{[1+(-1)^{l}]}{2}f(k)\)\Delta_l(k)\Delta^*_{l}(k)\ .
%+9\pi\frac{[1+(-1)^{l}]}{2}\int \frac{dk}{k}f(k)\Delta_l(k)\Delta^*_{l}(k)\ .
\end{equation}
For simplicity, we can parameterize $f(k)$ to be a power-law function as follows
\begin{equation}
f(k)=a_f\left(\frac{k}{k_0}\right)^{b_f}\ ,
\end{equation}
where $a_f$ and $b_f$ denote the magnitude and the spectral index, respectively.
Therefore, we can rewrite the angular power spectrum of CMB temperature anisotropies as
\begin{eqnarray}
\nonumber C_{l}&=&4\pi\int\frac{dk}{k}P_R(k)\left[1+B_0\left(\frac{k}{k_0}\right)^{-\alpha}\frac{[1+(-1)^{l}]}{2}\right]\Delta_l(k)\Delta^*_{l}(k),\\
\end{eqnarray}
where we define two new parameters, namely, $B_0=a_f/A_s$ and $\alpha=n_s-b_f-1$. In the next section, we will infer the constraints on these parameters through analyzing the low-multipole angular power spectrum of CMB temperature anisotropies in the PLANCK data.

\section{Data analysis and Results}\label{sec:result}\noindent
We fit the two parameters $B_0$ and $\alpha$ to the low-multipole (2$\leqslant l \leqslant$ 29) TT angular power spectrum in the PLANCK CMB data.
To describe the parity asymmetry, we use an estimator g($l$) %give the reference here from where you have taken this
which is defined as
\begin{equation}
%g(l)=\frac{\sum\limits_{l_0=2}^{l}l(l+1)C^{+}_l}{\sum\limits_{l_0=2}^{l}l(l+1)C^{-}_l},
g(l)=\frac{\sum\limits_{l}^{l+2}l(l+1)C^{+}_l}{\sum\limits_{l+1}^{l+3}l(l+1)C^{-}_l}\ ,
\end{equation}
where $l=4n+2$ and $0\leqslant n\leqslant6$, and $C^{\pm}_l$ are defined as
\begin{eqnarray}
C^{+}_l=\frac{1+(-1)^l}{2}C_l\ ,~~~~
C^{-}_l=\frac{1+(-1)^{l+1}}{2}C_l\ .
\end{eqnarray}
To calculate the CMB angular power spectrum, the six base cosmological parameters (i.e., the baryon density today $[\Omega_b h^2]$, the cold dark matter density today $[\Omega_c h^2]$, the angular scale of the sound horizon at last-scattering $[\theta_{\rm MC}]$, the reionization optical depth $[\tau]$, the amplitude of scalar power spectrum $[A_s]$, and the spectral index of scalar power spectrum $[n_s]$) are fixed to be their central values ($\Omega_bh^2=0.02222$, $\Omega_ch^2=0.1197$, $\theta_{\rm MC}=0.0104085$, $\tau=0.078$, $A_s=2.195\times 10^{-9}$, $n_s=0.9655$), which were given by the PLANCK TT+lowP data \citep{Ade:2015xua}. %(see Table \ref{tab:cosmol_parameters}).
\iffalse\begin{table}[htbp]
\begin{tabular}{|c|c|}
  \hline
   Parameter & $PLANCK~TT+lowP$ \\
  \hline
  $\Omega_bh^2$ & $0.02222\pm 0.00023$ \\
  $\Omega_ch^2$ & $0.1197\pm 0.0022$ \\
  $100\theta_{MC}$ & $1.04085\pm 0.00047$ \\
  $\tau$ & $0.078\pm 0.019$ \\
  $ln(10^{10}A_s)$ & $3.089\pm 0.036$ \\
  $n_s$ & $0.9655\pm 0.0062$ \\
  \hline
\end{tabular}
\caption{\small{The best-fit cosmological parameters and their $1\sigma$ uncertainties from  the PLANCK
TT+lowP likelihoods \cite{Ade:2015xua}.}}\label{tab:cosmol_parameters}
\end{table}\fi

To infer the parameter constraints in this work, we employ the $\chi^2$ statistic as follows
\begin{equation}
%\chi^2=\sum\limits_{l=3}^{29}\frac{(g^F(l)-g^P(l))^2}{\sigma_{P,l}^2}.
\chi^2=\sum\limits_{n=0}^{6}\frac{(g_n^F(l)-g_n^P(l))^2}{\sigma_{P,n}^2}\ ,
\end{equation}
where $g_n^F(l)$ and $g_n^P(l)$ are given by the Finslerian model and the PLANCK data, respectively. The $\sigma_{P,n}$ represents the uncertainty of $g^P_n(l)$.
Given the uncertainty $\sigma_l$ for each $C_l$, we use the formula of propagation of error to calculate the $\sigma_{P,n}^2$.
%, namely
%\begin{eqnarray}
%\sigma_{A+B}=\sqrt{\sigma_A^2+\sigma_B^2}\ ,\\
%\sigma_{\frac{A}{B}}=\left|\frac{A}{B}\right|\sqrt{(\frac{\sigma_A}{A})^2+(\frac{\sigma_B}{B})^2}\ .
%\end{eqnarray}
Due to difference between the upper deviation ${\sigma}_{l}^{\textrm{upper}}$ and lower deviation ${\sigma}_{l}^{\textrm{lower}}$ of $C_l$, we consider an approach to take account of $\sigma_{l}$ in our analysis, namely
\begin{equation}
\sigma_{l}=\sqrt{\frac{({\sigma}_{l}^{\textrm{upper}})^2+({\sigma}_{l}^{\textrm{lower}})^2}{2}}\ ,
%\sigma_{l}=\frac{{\sigma}_{l}^{\textrm{upper}}+{\sigma}_{l}^{\textrm{lower}}}{2}\ , ~~\sigma_{l}=\textrm{min}\{{\sigma}_{l}^{\textrm{upper}},~{\sigma}_{l}^{\textrm{lower}}\}\ , ~~\sigma_{l}=\textrm{max}\{{\sigma}_{l}^{\textrm{upper}},~{\sigma}_{l}^{\textrm{lower}}\}\ .
\end{equation}
which is used for the calculation of $\sigma_{P,n}$. %Correspondingly, the $\chi^2$ is calculated with the $\sigma_{P,n}$.
In our fitting, the prior probability distribution functions of the two Finslerian parameters are assumed to be uniform, namely, $B_0\in[-1,1]$ and $\alpha\in[0,5]$.

Our results are as follows. At $68\%$ confidence level, the constraints on the two Finslerian parameters are given by
\begin{eqnarray}
B_0&=&(-1.903\pm1.430)\times 10^{-3}\ ,\\\alpha&=&1.149^{+0.110}_{-0.284}\ .
\end{eqnarray}
Here $B_0$ is constrained to be of order $10^{-3}$, and deviates from zero by around $1.3$ standard deviation.
%just as mentioned above that the effect of $\textbf{b}(\textbf{k})$ is the first-order correction to the power spectrum in the $\Lambda$CDM model. %and $\alpha$ is constrained to be around 1.1.
\iffalse\begin{table}[htbp]
\begin{tabular}{|c|c|}
  \hline
  \multicolumn{1}{|l|}{Parameter}&  \\
  \hline
  \multicolumn{1}{|l|}{$B_0$} & $(-1.903\pm1.43)\times 10^{-3}$ \\
  \hline
  \multicolumn{1}{|l|}{$\alpha$} & $1.149^{+0.11}_{-0.284}$ \\
  \hline
\end{tabular}
\caption{\small{The best-fit values for the two Finslerian parameters and their 1$\sigma$ uncertainties from $\chi^2$ in the multipoles range $l=2-29$.}}\label{tab:bestfit results}
\end{table}\fi

\begin{figure}
\begin{center}
\includegraphics[width=0.8\columnwidth]{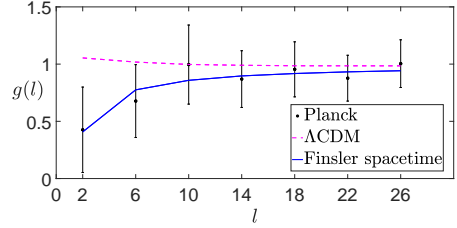}
\caption{The estimator g($l$) versus the seven multipole bins within 2$\leqslant l \leqslant$ 29. The solid line stands for g($l$) in the Finsler space-time. The dashed line represents g($l$) in the $\Lambda$CDM model. The black points represent $g_n^P(l)$ in PLANCK TT data, and the error bars represent the relating $1\sigma$ deviations.}
\label{figure1}
\end{center}
\end{figure}

Using the best-fit values of the Finslerian parameters, we depict the estimator $g(l)$ versus the seven multipole bins within $2\leq l \leq 29$ in Fig.~\ref{figure1}. The black points represent $g_n^P(l)$ obtained from PLANCK temperature data, and the error bars represent the relating $1\sigma$ deviations. The prediction of the Finslerian model, denoted by the solid line, is well consistent with the PLANCK temperature data within $1\sigma$ uncertainty.
%The values of g($l$) given by the Finslerian model are consistent with those obtained from PLANCK data within 1$\sigma$ uncertainty.
Therefore, we conclude that the Finslerian spacetime can account for the parity asymmetry in the CMB temperature anisotropies to some extent. In addition, we also depict the prediction of the best-fit $\Lambda$CDM model, denoted by the dashed line, for comparison.

\begin{figure}
\begin{center}
\includegraphics[width=0.8\columnwidth]{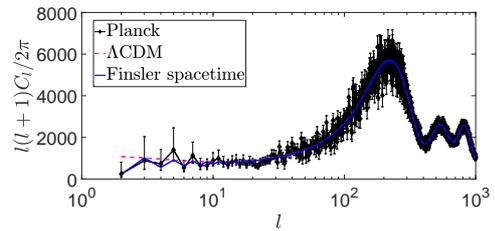}
\caption{The angular power spectrum of CMB temperature anisotropies. We use the best-fit values for the two Finslerian parameters. The solid line stands for $C_l$ in the Finsler space-time. The dashed line represents $C_l$ in the $\Lambda$CDM model. The line with diamonds represents the PLANCK 2015 temperature power spectrum, and the error bars show the relating 1$\sigma$ uncertainties.}
\label{figure2}
\end{center}
\end{figure}

Using the best-fit values of the Finslerian parameters, we also depict the angular power spectrum of CMB temperature anisotropies in Fig.~\ref{figure2}.
For comparison, we depict the PLANCK temperature data and the prediction of best-fit $\Lambda$CDM model.
We can see that the Finslerian model is well consistent with PLANCK temperature data within 1$\sigma$ uncertainty.
In addition, we infer that the low-$l$ power suppression of CMB temperature anisotropies can be also resolved in the Finslerian model.
%We infer that the power suppression of the CMB temperature anisotropies may relate to the parity asymmetry. In conclusion,
Therefore, we suggest that the parity asymmetry and the power deficit in the CMB temperature power spectrum may not be independent anomalies, and they may have a common origin.

\section{Conclusions}\label{sec:conclusion}\noindent
In this paper, we proposed to explain the parity asymmetry and the power deficit of CMB temperature anisotropies in the framework of Finsler space-time. We first reviewed the derivation of comoving curvature perturbation in Finslerian spacetime, and then calculated the angular power spectrum of temperature anisotropies in the CMB. We showed that the Finslerian correction to the CMB temperature fluctuations, \emph{i.e.,} the quadrupolar modulation, can account for the parity asymmetry phenomenologically. Finally, we estimated the cosmological constraints on the two Finslerian parameters using the CMB temperature power spectrum within 2$\leqslant l \leqslant$ 29 in PLANCK 2015 data. Our results showed that the Finslerian model can nicely account for the observed parity asymmetry of the CMB temperature anisotropies.
In addition, we noticed that the power of low-$l$ multipoles in the CMB temperature fluctuations is suppressed in the Finslerian model.
This power suppression predicted by the Finslerian model can account for the observed power deficit.
Therefore, this study suggested that the parity asymmetry and the power deficit of the CMB temperature anisotropies may stem from a common origin.
This conclusion is compatible with the existing one in ref. \citet{Kim:2010gf}, which showed that the low quadrupole anomaly could be related with the parity asymmetry.

\section*{Acknowledgments}\noindent
We appreciate Dr. S. Wang and H.-N. Lin for comprehensive comments on this work. DZ thanks Dr. S.-Y. Li, Y. Liu, and Y.-P. Li for fruitful discussions. This work has been funded by the National Natural Science Foundation of China under Grant no. 11675182, 11690022 and 11603005.

%%%%%%%%%%%%%%%%%%%%%%%%%%%%%%%%%%%%%%%%%%%%%%%%%%



%\bibliographystyle{mnras}

\bibliographystyle{mnras}
%\bibliography{ref}

\begin{thebibliography}{}
\makeatletter
\relax
\def\mn@urlcharsother{\let\do\@makeother \do\$\do\&\do\#\do\^\do\_\do\%\do\~}
\def\mn@doi{\begingroup\mn@urlcharsother \@ifnextchar [ {\mn@doi@}
  {\mn@doi@[]}}
\def\mn@doi@[#1]#2{\def\@tempa{#1}\ifx\@tempa\@empty \href
  {http://dx.doi.org/#2} {doi:#2}\else \href {http://dx.doi.org/#2} {#1}\fi
  \endgroup}
\def\mn@eprint#1#2{\mn@eprint@#1:#2::\@nil}
\def\mn@eprint@arXiv#1{\href {http://arxiv.org/abs/#1} {{\tt arXiv:#1}}}
\def\mn@eprint@dblp#1{\href {http://dblp.uni-trier.de/rec/bibtex/#1.xml}
  {dblp:#1}}
\def\mn@eprint@#1:#2:#3:#4\@nil{\def\@tempa {#1}\def\@tempb {#2}\def\@tempc
  {#3}\ifx \@tempc \@empty \let \@tempc \@tempb \let \@tempb \@tempa \fi \ifx
  \@tempb \@empty \def\@tempb {arXiv}\fi \@ifundefined
  {mn@eprint@\@tempb}{\@tempb:\@tempc}{\expandafter \expandafter \csname
  mn@eprint@\@tempb\endcsname \expandafter{\@tempc}}}

\bibitem[\protect\citeauthoryear{Abramo, Bernui, Ferreira, Villela  \&
  Wuensche}{Abramo et~al.}{2006}]{Abramo:2006gw}
Abramo L.~R.,  Bernui A.,  Ferreira I.~S.,  Villela T.,   Wuensche C.~A.,
  2006, \mn@doi [Phys. Rev. D] {10.1103/PhysRevD.74.063506}, 74, 063506

\bibitem[\protect\citeauthoryear{Adam et~al.}{Adam et~al.}{2016}]{Adam:2015rua}
Adam R.,  et~al., 2016, \mn@doi [Astron. Astrophys.]
  {10.1051/0004-6361/201527101}, 594, A1

\bibitem[\protect\citeauthoryear{Ade et~al.}{Ade et~al.}{2016a}]{Ade:2015xua}
Ade P. A.~R.,  et~al., 2016a, \mn@doi [Astron. Astrophys.]
  {10.1051/0004-6361/201525830}, 594, A13

\bibitem[\protect\citeauthoryear{Ade et~al.}{Ade et~al.}{2016b}]{Ade:2015hxq}
Ade P. A.~R.,  et~al., 2016b, \mn@doi [Astron. Astrophys.]
  {10.1051/0004-6361/201526681}, 594, A16

\bibitem[\protect\citeauthoryear{Aghanim et~al.}{Aghanim
  et~al.}{2016}]{Aghanim:2015xee}
Aghanim N.,  et~al., 2016, \mn@doi [Astron. Astrophys.]
  {10.1051/0004-6361/201526926}, 594, A11

\bibitem[\protect\citeauthoryear{Akbar-Zadeh}{Akbar-Zadeh}{1988}]{HA:z}
Akbar-Zadeh H.,  1988, Acad. Roy. Belg. Bull. Cl. Sci., 74, 281

\bibitem[\protect\citeauthoryear{Aluri \& Jain}{Aluri \&
  Jain}{2012}]{Aluri:2011wv}
Aluri P.~K.,  Jain P.,  2012, \mn@doi [Mon. Not. Roy. Astron. Soc.]
  {10.1111/j.1365-2966.2011.19981.x}, 419, 3378

\bibitem[\protect\citeauthoryear{Bao, Chern  \& Shen}{Bao
  et~al.}{2000}]{Chern:2000}
Bao D.,  Chern S.~S.,   Shen Z.,  2000, {An Introduction to Riemann Finsler
  Geometry , Graduate Texts in Mathematics 200}.
Springer, New York

\bibitem[\protect\citeauthoryear{Bennett et~al.}{Bennett
  et~al.}{2013}]{Bennett:2012zja}
Bennett C.~L.,  et~al., 2013, \mn@doi [Astrophys. J. Suppl.]
  {10.1088/0067-0049/208/2/20}, 208, 20

\bibitem[\protect\citeauthoryear{Bonga \& Gupt}{Bonga \&
  Gupt}{2016}]{Bonga:2015kaa}
Bonga B.,  Gupt B.,  2016, \mn@doi [Gen. Rel. Grav.]
  {10.1007/s10714-016-2071-0}, 48, 71

\bibitem[\protect\citeauthoryear{Cai, Ferreira, Hu  \& Quintin}{Cai
  et~al.}{2015}]{Cai:2015xla}
Cai Y.-F.,  Ferreira E. G.~M.,  Hu B.,   Quintin J.,  2015, \mn@doi [Phys. Rev.
  D] {10.1103/PhysRevD.92.121303}, 92, 121303

\bibitem[\protect\citeauthoryear{Chang \& Wang}{Chang \&
  Wang}{2013}]{Chang:2013vla}
Chang Z.,  Wang S.,  2013, \mn@doi [Eur. Phys. J. C]
  {10.1140/epjc/s10052-013-2516-5}, 73, 2516

\bibitem[\protect\citeauthoryear{Chang, Li, Li  \& Wang}{Chang
  et~al.}{2012}]{Chang:2011jw}
Chang Z.,  Li M.-H.,  Li X.,   Wang S.,  2012, \mn@doi [Eur. Phys. J. C]
  {10.1140/epjc/s10052-012-1915-3}, 72, 1915

\bibitem[\protect\citeauthoryear{Chang, Li  \& Wang}{Chang
  et~al.}{2015}]{Chang:2013lxa}
Chang Z.,  Li X.,   Wang S.,  2015, \mn@doi [Chin. Phys. C]
  {10.1088/1674-1137/39/5/055101}, 39, 055101

\bibitem[\protect\citeauthoryear{Chang, Rath, Sang  \& Zhao}{Chang
  et~al.}{2018}]{Chang:2018msh}
Chang Z.,  Rath P.~K.,  Sang Y.,   Zhao D.,  2018, \mn@doi [Res. Astron.
  Astrophys.] {10.1088/1674-4527/18/3/29}, 18, 029

\bibitem[\protect\citeauthoryear{Copi, Huterer  \& Starkman}{Copi
  et~al.}{2004}]{Copi:2003kt}
Copi C.~J.,  Huterer D.,   Starkman G.~D.,  2004, \mn@doi [Phys. Rev. D]
  {10.1103/PhysRevD.70.043515}, 70, 043515

\bibitem[\protect\citeauthoryear{Copi, Huterer, Schwarz  \& Starkman}{Copi
  et~al.}{2009}]{Copi:2008hw}
Copi C.~J.,  Huterer D.,  Schwarz D.~J.,   Starkman G.~D.,  2009, \mn@doi [Mon.
  Not. Roy. Astron. Soc.] {10.1111/j.1365-2966.2009.15270.x}, 399, 295

\bibitem[\protect\citeauthoryear{Copi, Huterer, Schwarz  \& Starkman}{Copi
  et~al.}{2015a}]{Copi:2013jna}
Copi C.~J.,  Huterer D.,  Schwarz D.~J.,   Starkman G.~D.,  2015a, \mn@doi
  [Mon. Not. Roy. Astron. Soc.] {10.1093/mnras/stv501}, 449, 3458

\bibitem[\protect\citeauthoryear{Copi, Huterer, Schwarz  \& Starkman}{Copi
  et~al.}{2015b}]{Copi:2013cya}
Copi C.~J.,  Huterer D.,  Schwarz D.~J.,   Starkman G.~D.,  2015b, \mn@doi
  [Mon. Not. Roy. Astron. Soc.] {10.1093/mnras/stv1143}, 451, 2978

\bibitem[\protect\citeauthoryear{Efstathiou}{Efstathiou}{2003}]{Efstathiou:2003hk}
Efstathiou G.,  2003, \mn@doi [Mon. Not. Roy. Astron. Soc.]
  {10.1046/j.1365-8711.2003.06940.x}, 343, L95

\bibitem[\protect\citeauthoryear{Eriksen, Banday, Gorski, Hansen  \&
  Lilje}{Eriksen et~al.}{2007}]{Eriksen:2007pc}
Eriksen H.~K.,  Banday A.~J.,  Gorski K.~M.,  Hansen F.~K.,   Lilje P.~B.,
  2007, \mn@doi [Astrophys. J.] {10.1086/518091}, 660, L81

\bibitem[\protect\citeauthoryear{Gruppuso, Finelli, Natoli, Paci, Cabella,
  De~Rosa  \& Mandolesi}{Gruppuso et~al.}{2011}]{Gruppuso:2010nd}
Gruppuso A.,  Finelli F.,  Natoli P.,  Paci F.,  Cabella P.,  De~Rosa A.,
  Mandolesi N.,  2011, \mn@doi [Mon. Not. Roy. Astron. Soc.]
  {10.1111/j.1365-2966.2010.17773.x}, 411, 1445

\bibitem[\protect\citeauthoryear{Hansen, Kim, Frejsel, Ramazanov, Naselsky,
  Zhao  \& Burigana}{Hansen et~al.}{2012}]{Hansen:2012qd}
Hansen M.,  Kim J.,  Frejsel A.~M.,  Ramazanov S.,  Naselsky P.,  Zhao W.,
  Burigana C.,  2012, \mn@doi [JCAP] {10.1088/1475-7516/2012/10/059}, 1210, 059

\bibitem[\protect\citeauthoryear{Hinshaw, Banday, Bennett, Gorski, Kogut, Smoot
   \& Wright}{Hinshaw et~al.}{1996}]{Hinshaw:1996uq}
Hinshaw G.,  Banday A.~J.,  Bennett C.~L.,  Gorski K.~M.,  Kogut A.,  Smoot
  G.~F.,   Wright E.~L.,  1996, \mn@doi [Astrophys. J.] {10.1086/310074}, 464,
  L17

\bibitem[\protect\citeauthoryear{Hinshaw et~al.}{Hinshaw
  et~al.}{2013}]{Hinshaw:2012aka}
Hinshaw G.,  et~al., 2013, \mn@doi [Astrophys. J. Suppl.]
  {10.1088/0067-0049/208/2/19}, 208, 19

\bibitem[\protect\citeauthoryear{Hsiang, Wu  \& Ford}{Hsiang
  et~al.}{2011}]{Hsiang:2010uv}
Hsiang J.-T.,  Wu C.-H.,   Ford L.~H.,  2011, \mn@doi [Phys. Lett. A]
  {10.1016/j.physleta.2011.04.052}, 375, 2296

\bibitem[\protect\citeauthoryear{Jain \& Rath}{Jain \&
  Rath}{2015}]{Jain:2014cpa}
Jain P.,  Rath P.~K.,  2015, \mn@doi [Eur. Phys. J. C]
  {10.1140/epjc/s10052-015-3333-9}, 75, 113

\bibitem[\protect\citeauthoryear{Kim \& Naselsky}{Kim \&
  Naselsky}{2010a}]{Kim:2010gd}
Kim J.,  Naselsky P.,  2010a, \mn@doi [Phys. Rev. D]
  {10.1103/PhysRevD.82.063002}, 82, 063002

\bibitem[\protect\citeauthoryear{Kim \& Naselsky}{Kim \&
  Naselsky}{2010b}]{Kim:2010gf}
Kim J.,  Naselsky P.,  2010b, \mn@doi [Astrophys. J.]
  {10.1088/2041-8205/714/2/L265}, 714, L265

\bibitem[\protect\citeauthoryear{Land \& Magueijo}{Land \&
  Magueijo}{2005a}]{Land:2005jq}
Land K.,  Magueijo J.,  2005a, \mn@doi [Phys. Rev. D]
  {10.1103/PhysRevD.72.101302}, 72, 101302

\bibitem[\protect\citeauthoryear{Land \& Magueijo}{Land \&
  Magueijo}{2005b}]{Land:2005ad}
Land K.,  Magueijo J.,  2005b, \mn@doi [Phys. Rev. Lett.]
  {10.1103/PhysRevLett.95.071301}, 95, 071301

\bibitem[\protect\citeauthoryear{Li \& Chang}{Li \& Chang}{2012}]{Li:2010wv}
Li X.,  Chang Z.,  2012, \mn@doi [Differ. Geom. Appl.]
  {10.1016/j.difgeo.2012.07.009}, 30, 737

\bibitem[\protect\citeauthoryear{Li \& Chang}{Li \& Chang}{2014}]{Li:2014taa}
Li X.,  Chang Z.,  2014, \mn@doi [Phys. Rev. D] {10.1103/PhysRevD.90.064049},
  90, 064049

\bibitem[\protect\citeauthoryear{Li \& Lin}{Li \& Lin}{2017}]{Li:2017vuc}
Li X.,  Lin H.-N.,  2017, \mn@doi [Eur. Phys. J. C]
  {10.1140/epjc/s10052-017-4897-3}, 77, 316

\bibitem[\protect\citeauthoryear{Li \& Wang}{Li \& Wang}{2016}]{Li:2015kua}
Li X.,  Wang S.,  2016, \mn@doi [Eur. Phys. J. C]
  {10.1140/epjc/s10052-016-3890-6}, 76, 51

\bibitem[\protect\citeauthoryear{Li, Lin, Wang  \& Chang}{Li
  et~al.}{2015a}]{Li:2015uda}
Li X.,  Lin H.-N.,  Wang S.,   Chang Z.,  2015a, \mn@doi [Eur. Phys. J. C]
  {10.1140/epjc/s10052-015-3380-2}, 75, 181

\bibitem[\protect\citeauthoryear{Li, Wang  \& Chang}{Li
  et~al.}{2015b}]{Li:2015sja}
Li X.,  Wang S.,   Chang Z.,  2015b, \mn@doi [Eur. Phys. J. C]
  {10.1140/epjc/s10052-015-3468-8}, 75, 260

\bibitem[\protect\citeauthoryear{Liu, Frejsel  \& Naselsky}{Liu
  et~al.}{2013}]{Liu:2013wfa}
Liu H.,  Frejsel A.~M.,   Naselsky P.,  2013, \mn@doi [JCAP]
  {10.1088/1475-7516/2013/07/032}, 1307, 032

\bibitem[\protect\citeauthoryear{Randers}{Randers}{1941}]{Randers:1941gge}
Randers G.,  1941, \mn@doi [Phys. Rev.] {10.1103/PhysRev.59.195}, 59, 195

\bibitem[\protect\citeauthoryear{Rath \& Jain}{Rath \&
  Jain}{2013}]{Rath:2013yra}
Rath P.~K.,  Jain P.,  2013, \mn@doi [JCAP] {10.1088/1475-7516/2013/12/014},
  1312, 014

\bibitem[\protect\citeauthoryear{Rath, Aluri  \& Jain}{Rath
  et~al.}{2015}]{Rath:2014cka}
Rath P.~K.,  Aluri P.~K.,   Jain P.,  2015, \mn@doi [Phys. Rev. D]
  {10.1103/PhysRevD.91.023515}, 91, 023515

\bibitem[\protect\citeauthoryear{Riotto}{Riotto}{2003}]{Riotto:2002yw}
Riotto A.,  2003, ICTP Lect. Notes Ser., 14, 317

\bibitem[\protect\citeauthoryear{Spergel et~al.}{Spergel
  et~al.}{2003}]{Spergel:2003cb}
Spergel D.~N.,  et~al., 2003, \mn@doi [Astrophys. J. Suppl.] {10.1086/377226},
  148, 175

\bibitem[\protect\citeauthoryear{Zhao}{Zhao}{2014}]{Zhao:2013jya}
Zhao W.,  2014, \mn@doi [Phys. Rev.] {10.1103/PhysRevD.89.023010}, D89, 023010

\bibitem[\protect\citeauthoryear{Zhao, Li, Wang  \& Chang}{Zhao
  et~al.}{2015}]{Zhao:2014kca}
Zhao D.,  Li M.-H.,  Wang P.,   Chang Z.,  2015, \mn@doi [Chin. Phys. C]
  {10.1088/1674-1137/39/9/095101}, 39, 095101

\makeatother
\end{thebibliography}

%%%%%%%%%%%%%%%%%%%%%%%%%%%%%%%%%%%%%%%%%%%%%%%%%%

% Don't change these lines
\bsp	% typesetting comment
\label{lastpage}
\end{document}